\documentclass[doublecol,a4paper]{epl2}
\usepackage{amsmath} 
	
\newcommand{\ee}{\text{e}}

\newcommand{\kb}{k_\text{\tiny B}}

\newcommand{\va}{v_\text{\tiny A}}
\newcommand{\fa}{f_\text{\tiny A}}

\newcommand{\td}{\tau_\text{r}}

\newcommand{\Ta}{T_\text{\tiny A}}

\newcommand{\g}{\gamma}

\newcommand{\xT}{x_\text{\tiny T}}
\newcommand{\xA}{x_\text{\tiny A}}

\graphicspath{{figures/}}

\providecommand{\avg}[1]{\left \langle #1 \right \rangle}
\providecommand{\pnt}[1]{\left ( #1 \right)}
\providecommand{\brt}[1]{\left [ #1 \right]}
\providecommand{\abs}[1]{\left | #1 \right|}
\providecommand{\f}[2]{\frac{ #1}{#2}}

\title{Activity driven fluctuations in living cells}

\author{\'E. Fodor\inst{1,*} \and M. Guo\inst{2,*} \and N. S. Gov\inst{3} \and
  P. Visco\thanks{Correspondig Author:
\email{paolo.visco@univ-paris-diderot.fr}}\inst{1}  \and D. A. Weitz\inst{2}  \and F. van Wijland \inst{1}}
\shortauthor{\'E. Fodor \etal}

\institute{                    
  \inst{1} Laboratoire Mati\`ere et Syst\`emes
  Complexes, UMR 7057 CNRS/P7, Universit\'e Paris Diderot, 10 rue
  Alice Domon et L\'eonie Duquet, 75205 Paris cedex 13, France \\
  \inst{2} School of Engineering and Applied
  Sciences, Harvard University, Cambridge MA 02138, USA\\
 \inst{3} Department of Chemical Physics,
  Weizmann Institute of Science, 76100 Rehovot, Israel\\
\inst{*} These authors contributed equally to this work
}

\abstract{ We propose a model for the dynamics of a probe embedded in
  a living cell, where both thermal fluctuations and nonequilibrium
  activity coexist. The model is based on a confining harmonic
  potential describing the elastic cytoskeletal matrix, which
  undergoes random active hops as a result of the nonequilibrium
  rearrangements within the cell.  We describe the probe's statistics
  and we bring forth quantities affected by the nonequilibrium
  activity. We find an excellent agreement between the predictions of
  our model and experimental results for tracers inside living
  cells. Finally, we exploit our model to arrive at quantitative
  predictions for the parameters characterizing nonequilibrium
  activity, such as the typical time scale of the activity and the
  amplitude of the active fluctuations.}

\begin{document}

\maketitle

Actin filaments are involved in a number of functions including cell
motility, adhesion, gene expression, and signalling. When fueled by
ATP supply, Myosin motors advance along these filaments by performing
a directed stochastic motion.  By tracking the trajectory of a
micron-size probe embedded within the cytoskeletal network, and by
subjecting it to microrheology experiments, one can hope to access and
understand some of the properties of the nonequilibrium activity
inside the cytoskeletal network.  Experiments were first carried out
in actin gels without molecular motors, known as {\it passive
  gels}~\cite{hou_brownian_1990,jones_tracer_1996,doi:10.1021/jp9072153,
  Apgar20001095, Tseng2002210}. Some progress in the experimental
field has provided new results for tracers attached to the cortex of
living cells~\cite{Bursac}, and also for {\it in vitro} actin
gels~\cite{toyota,stuhrmann}. In such gels, called {\it active gels},
the tracer dynamics exhibits large excursions corresponding to
directed motion events, in addition to the thermal fluctuations
already observed in passive gels. Due to the active processes, the
actin network fluctuations comprise a strongly nonequilibrium
component. Experimentally, the out-of-equilibrium nature of such
activity has been evidenced by the violation of the fluctuation
dissipation theorem (FDT)~\cite{Mizuno,Betz,gallet09}. To account for
nonequilibrium activity, a generalization of the FDT has been
developed introducing a frequency dependent effective
temperature~\cite{Nir,Joanny,Loi}. This generalization is based on a
description of tracers dynamics at a mesoscopic scale, which can be
described using a generalized Langevin
equation~\cite{Lau,Bohec,Mason}. At a macroscopic scale, the dynamics
of acto-myosin networks have been described \textit{via} hydrodynamic
treatments~\cite{Prost} or polymer
theory~\cite{mackintosh1,mackintosh2}.

In what follows, we present results of microrheology experiments in
the cytoplasm of living cells, which are characterized by a highly
nonequilibrium activity. Along with experiments, we propose a model
which mixes simple but nontrivial rheology with random fluctuations
due to active processes inside the cell. We carry out a comparison
with experimental data, which allows us to directly determine some
microscopic mechanisms that drive active fluctuations inside the
cell. We demonstrate that our quantitive estimation of the
nonequilibrium active features is consistent with different kinds of
experimental measurements, thus supporting the overall consistency of
our model.

\begin{figure}
\includegraphics[width=0.95 \columnwidth]{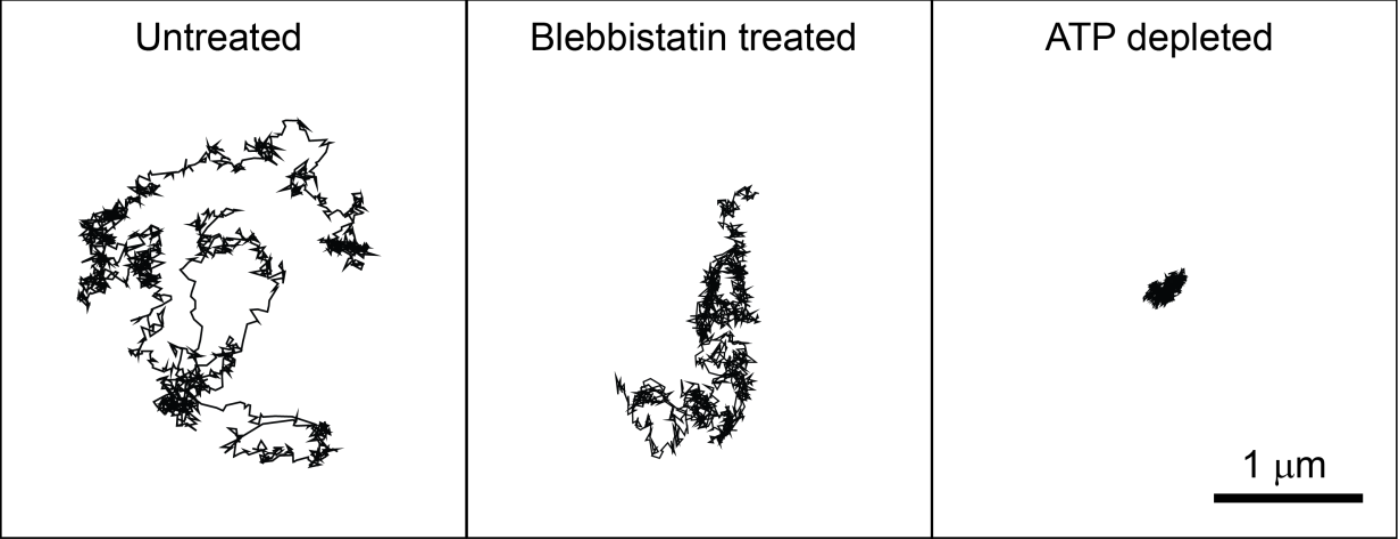}
\vskip.3cm
\includegraphics[width=0.95 \columnwidth]{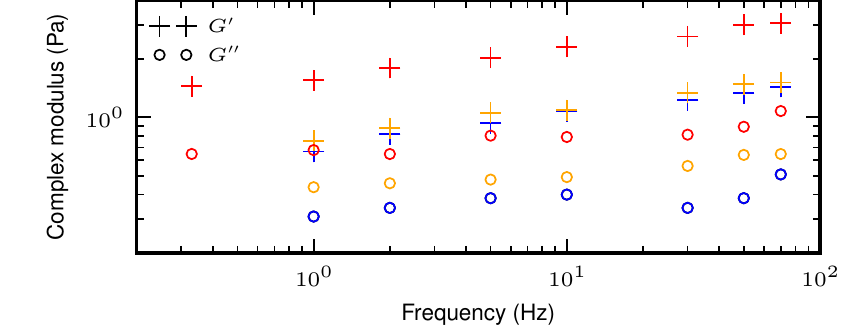}
\caption{\label{fig:traj}
(Top)~Typical trajectories of $200$~nm PEG coated
  beads in A7 cells under three conditions: control, $10$~$\mu$M blebbistatin
  treatment, ATP depletion. Trajectory length is about $2$~min.
(Bottom)~Elastic storage modulus $G'$ ($+$) and loss
  modulus $G''$ ($\circ$) from active microrheology experiments in
  untreated (red), blebbistatin treated (orange), and ATP depleted
  (blue) A7 cells.}
\end{figure}

\begin{figure}
\includegraphics[width=0.95 \columnwidth]{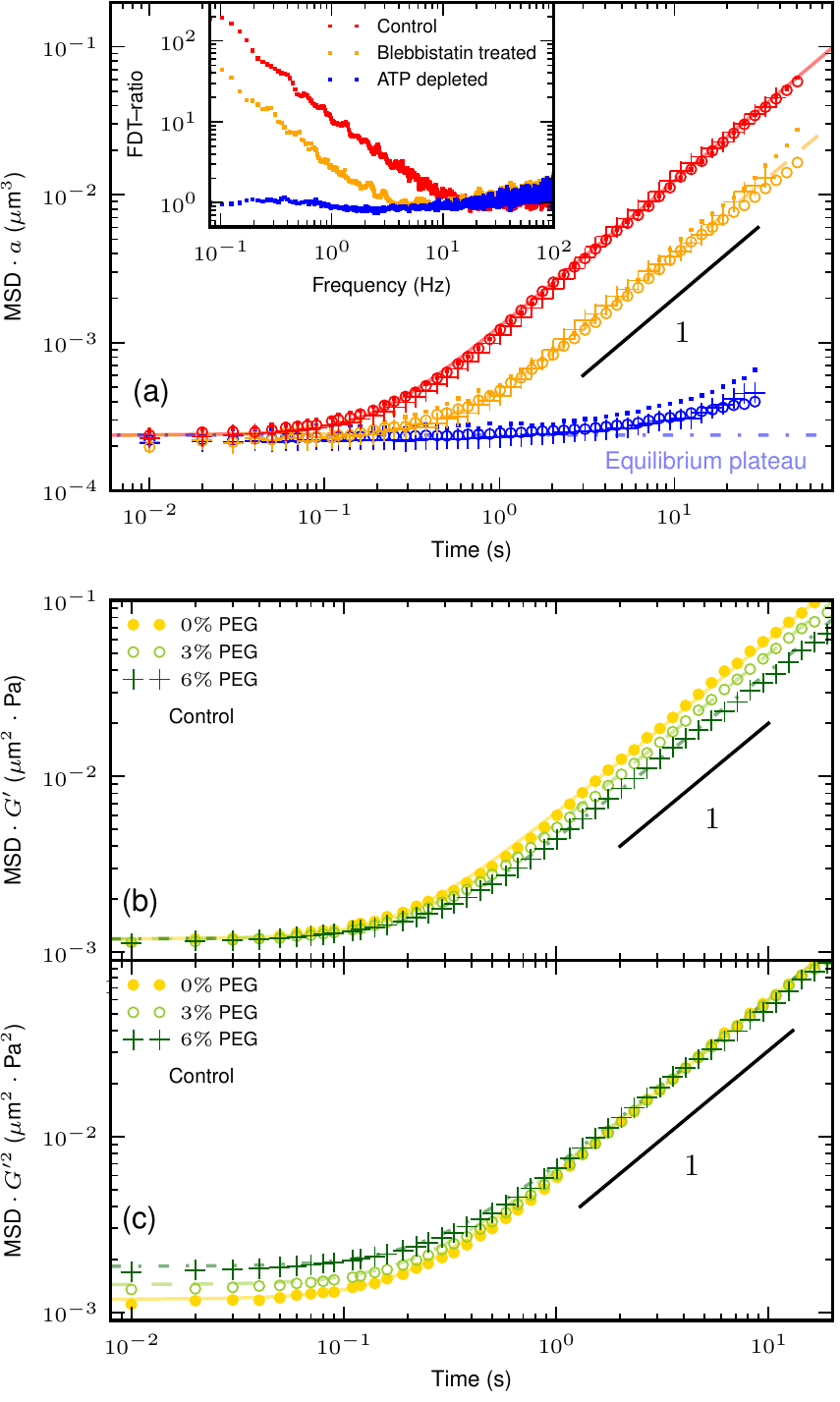}
\caption{\label{fig:plot}(a)~Time evolution of the
  one-dimensional mean square displacement scaled with the tracer
  radius $a=50$ ($+$), $100$ ($\circ$) and $250$~nm ($\cdot$) for
  control (red), blebbistatin treated (orange), and ATP depleted
  (blue) cells, and the corresponding best fitting curves
  (Eq.~\eqref{eq:msd}): solid, dashed, and dot dashed line,
  respectively.  (Inset)~FDT-ratio as a function of frequency. It
  equals $1$ in ATP depleted cells as for an equilibrium system, and
  it deviates from it in the two other conditions at small frequency
  showing that nonequilibrium processes drive the dynamics in this
  regime.  (b)~Time evolution of the MSD scaled with $G'$ measured
  with tracers of radius $a=100$~nm in control cells. The $G'$ value
  increases with the percentage of PEG introduced in the cell: $0\%$
  (yellow $\bullet$), $3\%$ (light green $\circ$), and $6\%$ (dark
  green $+$). The best fit curves are in solid, dashed, and dot dashed
  lines, respectively.  The short time scale plateau scales like
  $1/G'$.  (c)~Time evolution of the MSD times $G'^2$. The large time
  diffusive part scales as $1/G'^2$.  }
\end{figure}

We inject sub-micron colloidal tracers in the cytoplasm of living A7
cells, and track a two-dimensional projection of their
  fluctuating (3-D) motion with confocal
microscopy~\cite{cunningham}. We observe some directed motion events
in the tracers' trajectories in addition to the thermal fluctuations
of small amplitudes (Fig.~\ref{fig:traj}), as already reported in
synthetic active gels~\cite{toyota}. To investigate the role of
biological activity in the intracellular mechanics, we subject cells
to two treatments. We inhibit Myosin II motors by adding 10~$\mu$M of
blebbistatin to the culture medium, and we deplete cells of ATP
through addition of $2$~mM sodium azide and $10$~mM of
2-deoxyglucose. We extract the one-dimensional mean square
displacement (MSD) from the spontaneous motion of tracers for
different radius sizes $a=\{50,100,250\}$~nm. We present the MSD
multiplied by $a$ for the control, blebbistatin and ATP depleted
conditions in Fig.~\ref{fig:plot}(a), showing that the MSD scales like
$1/a$. The small time MSD is constant in the three conditions, while
the large time behavior is diffusive, apart for ATP depleted cells,
where it remains almost constant. Since the time evolution of the MSD
is qualitatively similar for tracers of different sizes, we deduce
that the tracers are bigger than the mesh size of the cytoskeletal
network, thus allowing us to consider that they evolve in a continuous
medium in first approximation.

We measure the mechanical properties of the cytoplasm \textit{via}
active microrheology method by using optical tweezers~\cite{guo13}. We
impose a sinusoidal oscillation on a particle with diameter
$0.5$~$\mu$m within the cytoplasm. From the resultant displacement of
the bead, we extract the complex modulus $G^*=1/(6\pi a \chi)$,
where $\chi$ is the Fourier response function. It reveals that it
weakly depends on frequency, and that the elastic contribution is
significantly larger than the dissipative one (Fig.~\ref{fig:traj}),
in agreement with previous results~\cite{fabry,guo13}. Moreover, we do
not observe a significant change in the cytoplasmic mechanical
property according to active processes. The cytoplasm is still mainly
elastic in blebbistatin treated and ATP depleted cells, with a storage
modulus being twice as small as in untreated cells where it equals
approximately $2$~Pa.

To quantify departure from equilibrium, we extract the FDT-ratio which
compares the active microrheology measurement with the random
intracellular motion visualized by tracer
particles~\cite{gallet09,Betz,Mizuno}. It is defined in terms of the
position power spectrum $\tilde C$ and the imaginary part of the
Fourier response function $\chi''$ as
$\text{FDT-ratio}(\omega)=-\omega\tilde
C(\omega)/\brt{2\chi''(\omega)\kb T}$, where $T$ is the bath
temperature.  It equals $1$ for an equilibrium system, and deviates
from it otherwise.  The control and blebbistatin treated cells are
out-of-equilibrium, whereas the effect of the nonequilibrium processes
are negligible in ATP depleted cells (Inset in
Fig.~\ref{fig:plot}(a)). This supports that the nonequilibrium
processes hibernate in the latter as long as no ATP supply is
provided, suggesting that there is an equilibrium reference state
where the tracer particle is trapped in an elastic cytoskeletal
network. Given we can not rely on equilibrium physics to describe the
tracer's dynamics in the two other conditions, we offer a new model to
characterize its nonequilibrium properties.

We vary experimentally the elastic modulus $G'$ by adding various
amount of $300$-Dalton polyethylene glycol (PEG) into the cell culture
medium~\footnote{After the stress, cells are allowed to equilibrate
  for $10$~min at $37^\circ$C and $5\%$ CO$_2$, before we perform the
  imaging or optical-tweezer measurement. The cell size and mechanics
  equilibrate in $2$~min after adding PEG based on our imaging and
  previous studies~\cite{Fredberg}.}. This results in an osmotic
compression on the cell, so that $G'$ increases with the amount of PEG
applied~\cite{Fredberg}. We report in Figs.~\ref{fig:plot}~(b)-(c) the
MSD data multiplied by $G'$ and $G'^2$ for different values of
$G'$. It appears the value of the small time plateau scales as $1/G'$
while the large time diffusion constant scales as $1/G'^2$.

\begin{figure}
\includegraphics[width=\columnwidth]{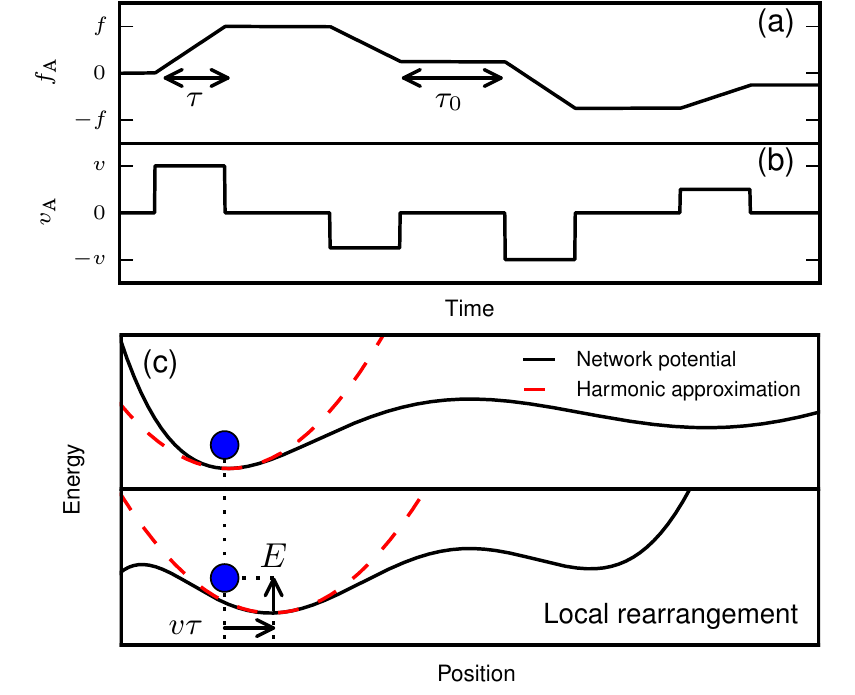}
\caption{\label{fig:act} Typical realization of (a)~the active force
  $\fa$, and (b)~the corresponding active bursts $\va$. $\fa$ is
  constant over a quiescence time of typical value $\tau_0$, and
  varies linearly with a slope uniformly distributed in $\brt{-f,f}$
  during a persistence time of order $\tau$. $\va$ is proportional to
  the time derivative of $\fa$.  (c)~Schematic representation of the
  energetic landscape rearrangement due to nonequilibrium activity and
  its modeling using the active burst applied on the local minimum. We
  depict the network potential in black solid line, the harmonic
  approximation in dashed red line, and the tracer particle in filled
  blue circle.  Nonequilibrium activity leads to a displacement
  $v\tau$ of the potential, resulting in an energy gain $E\simeq
  k(v\tau)^2$ for the tracer.}
\end{figure}

The cytoskeleton acts as a thermostat for the tracer
particle. Provided that inertial effects are negligible in the
intracellular environment, we model the dynamics of the tracer's
position ${\bf r}$ by means of an overdamped Langevin equation. We use
a harmonic approximation to account for the interaction of the tracer
with the surrounding network. The main new ingredient of our model
lies in expressing the effect of nonequilibrium activity. We postulate
that the underlying action of the active processes induces local
rearrangements of the network, resulting in an active force applied on
the tracers. As an example of such nonequilibrium processes, the
activity of Myosin II motors can slide cytoskeletal filaments past
each others leading to a local deformation of the
network~\cite{toyota}. To account for the directed motion events
observed in our experimental trajectories, we consider that the active
force proceeds by a sequence of rapid ballistic jumps followed by
quiescent periods. It remains constant during intervals of average
quiescence time $\tau_0$, when the tracer is only subjected to thermal
fluctuations, and it varies during a persistence time of order $\tau$
by a quantity ${\bf\fa}=f\hat{\bf n}$, where $\hat{\bf n}$ is a random
direction in the three-dimensional space. We assume that the
persistence and quiescence times are exponentially distributed
variables as observed in synthetic active
gels~\cite{toyota,stuhrmann,silva}, and that they do not depend on the
network and tracer properties. Putting these ingredients together, we
arrive at the equation for $x$, the one-dimensional projection of
${\bf r}$
\begin{equation}\label{eq-model}
\gamma\frac{d x}{d t}=-k x+\xi+\fa
\,,
\end{equation}
where $\xi$ is a zero mean Gaussian white noise with correlations
$\langle \xi(t)\xi(t')\rangle=2\gamma\kb T\delta(t-t')$, and $\fa$ is
a random force with typical realization described in
Fig.~\ref{fig:act}(a). The spring constant of the surrounding network
is $k$, and $\gamma$ is the friction coefficient of the
environment. Our model is associated with a Fourier response
  function $\chi=1/(k + i \omega \gamma)$, from which we deduce that
  complex modulus is of the form $G^*=1/(6\pi a \chi)=k/(6\pi
  a)+i\omega\eta$, where $\eta$ is the viscosity of the fluid
surrounding the tracer~\cite{mason00,Mason}. We neglect the weak
frequency dependence of the real part $G'$ as determined from active
microrheology measurements, so that the spring constant is directly
given by $k=6\pi a G'$, as already reported in other complex
  fluids with similar elastic behavior~\cite{mason00}. Stokes' law
ensures that $\gamma$ is independent of $G'$, and $\gamma\propto a$.

To illustrate our model with an immediate physical picture, we
introduce the variable ${\bf r}_0={\bf \fa}/k$ which we regard as the
position of the local minimum of the potential in which the tracer is
trapped. The local rearrangements of the network due to nonequilibrium
activity result in a shift of the local minimum the tracer sits
in. Thus, this position has a dynamics of its own given by a random
active burst ${\bf \va}$ in which a burst $v\hat{\bf n}$ is felt
during the persistence time, while it equals zero during the
quiescence time (Fig.~\ref{fig:act}(b)). The active force projection
is simply related to the active burst projection as $d \fa/d
t=k\va$. We assume that the typical variation $f$ of the active force
is independent of the network properties, whereas the active burst
amplitude $v=f/(k\tau)$ depends on the properties of the cytoskeletal
network \textit{via} $k$.

From the Fourier transform of Eq.~\eqref{eq-model}, we compute the
position autocorrelation function $C(t)=\avg{x(t)x(0)}$, and then
deduce the one-dimensional MSD as $\avg{\Delta x^2}(t)=2(C(0)-C(t))$.
We denote the thermal contribution to the MSD by $\avg{\Delta \xT^2}$,
and the MSD when the particle is only subjected to motor activity by
$\avg{\Delta \xA^2}$, so that: $\avg{\Delta x^2}=\avg{\Delta
  \xT^2}+\avg{\Delta \xA^2}$. The thermal MSD is the same as for the
Ornstein-Uhlenbeck process, and we compute the active contribution in
terms of the parameters characterizing the active force:
\begin{subequations}
\label{eq:msd}
\begin{eqnarray}
\avg{\Delta x_\text{\tiny T}^2}(t) &=& \f{2\kb T}{k}\pnt{1-\ee^{-t/\td}}
\,,
\\
\avg{\Delta x_\text{\tiny A}^2}(t) &=& \f{2 \kb \Ta/k}{1-(\tau/\td)^2} \bigg[ \pnt{\f{\tau}{\td}}^3 \pnt{ 1-\ee^{-t/\tau} -\f{t}{\tau} }
\nonumber
\\
& &+ \ee^{-t/\td} +\f{t}{\td} -1 \bigg]
\,,
\end{eqnarray}
\end{subequations}
where $\td=\g/k$ is a microscopic relaxation time scale.  In the
passive case, \textit{i.e.} when $\Ta=0$, it saturates to the value $2
\kb T/k$ within a time $\td$ as predicted by the equipartition
theorem, meaning that the tracer is confined in the cytoskeleton. The
active force represents the random fluctuations of the cytoskeletal
network induced by the nonequilibrium activity. With such a force, the
MSD exhibits a plateau at the equilibrium value corresponding to a
transient elastic confinement at times $\td\ll t\ll\tau$, and then has
a diffusion-like growth on longer times with coefficient $2 \kb \Ta
/\g$. Provided that $k\propto G'$, it follows that the equilibrium
plateau scales like $1/G'$, as we observe experimentally
(Fig.~\ref{fig:plot}(b)). The energy scale $\kb \Ta=\g (v\tau)^2 /[3(
  \tau+\tau_0)]$ defines an active temperature, which is related to
the amplitude of the active fluctuations as defined by the active
burst correlations $\avg{\va (t) \va(0)}=\kb\Ta
\ee^{-|t|/\tau}/(\tau\g)$. The independence of $f$ and $\tau$ with
respect to $G'$ yields $v\propto1/G'$, from which we deduce that the
large time diffusion coefficient scales as $1/G'^2$, in agreement with
our measurements (Fig.~\ref{fig:plot}(c)).

On the basis of our phenomenological picture where the nonequilibrium
dynamics is driven by an active remodelling of the cytoskeletal
network, we propose a physical argument for the scaling of the MSD
with the tracers' size $a$ presented in Fig.~\ref{fig:plot}(a). As
presented above, we first assume that $k$ and $\gamma$ scales like
$a$. Within our model, the active burst represents the activity-driven
network deformation and reorganization, which result in a change of
the tracer's local energetic landscape. During a burst event, the
local minimum is shifted by a random amount. Regarding this event as
instantaneous, the tracer finds itself at a distance of order $v \tau$
from the new local minimum position after each burst. It follows that
the typical energy provided by nonequilibrium activity to the particle
is $E\simeq k (v \tau)^2$, as depicted in Fig.~\ref{fig:act}(c). We
assume that it does not depend on the particle properties, just as
$\tau$ and $\tau_0$, thus being independent of the tracer's typical
size $a$. Since $k\propto a$, we deduce $v\propto1/\sqrt{a}$, implying
that $\Ta$ is independent of $a$. Finally, the relaxation time $\td$
is also independent of $a$, leading to a scaling of the MSD like $1/a$
which agrees with our observation.

We use our analytic expression to fit the MSD data multiplied by $a$
for the three conditions described above. We assume the viscosity of
the fluid surrounding the tracer is the cytoplasm viscosity
$\eta\sim10^{-3}$~Pa$\,\cdot\,$s~\cite{Mastro}, and we deduce the
damping coefficient from Stokes' law: $\gamma=6\pi a \eta$. We
estimate the $k$ value from the small time plateau.  The only
remaining parameters are the ones characterizing nonequilibrium
activity: $\Ta/T=\{2.8,0.9\}\times10^{-3}$,
$\tau=\{0.16\pm0.03,0.39\pm0.09\}$~s, in control and blebbistatin
treated cells, respectively. The estimation error made on $\Ta/T$ is
of the order of $1\%$ in control, and $0.1\%$ in blebbistatin treated
cells.

The amplitude of the active fluctuations is smaller in blebbistatin
treated cells, meaning that the inhibition of Myosin II motors reduces
the proportion of nonequilibrium fluctuations with respect to the
thermal ones as expected. Other nonequilibrium processes drive the
out-of-equilibrium dynamics in this condition. The typical time scale
$\tau$ of the persistent motion events is enhanced in blebbistatin
treated cells. Assuming that each active burst persists until the
stress that accumulates in the network causes the network to locally
fail, weaker motors due to the addition of blebistatin will contract
for a longer duration until such a critical stress builds up. Provided
that $1/\tau$ is the typical frequency below which the nonequilibrium
processes affect the dynamics, this supports that the active
fluctuations take over the thermal ones at larger frequencies in the
control cells compared with the blebbistatin treated ones. Notice that
$\Ta$ represents the ability of the tracer to diffuse on long times,
and $T$ quantifies here only the motion of the bead at short times
when it is trapped within the elastic cytoskeletal network. The fact
that we find $\Ta$ small compared to $T$ does not mean that the active
processes are negligible, as they control entirely the long-time and
long-distance diffusion of the tracer. In the absence of activity, the
tracer does not diffuse at all and remains trapped in the elastic
network.

\begin{figure}
\includegraphics[width=\columnwidth]{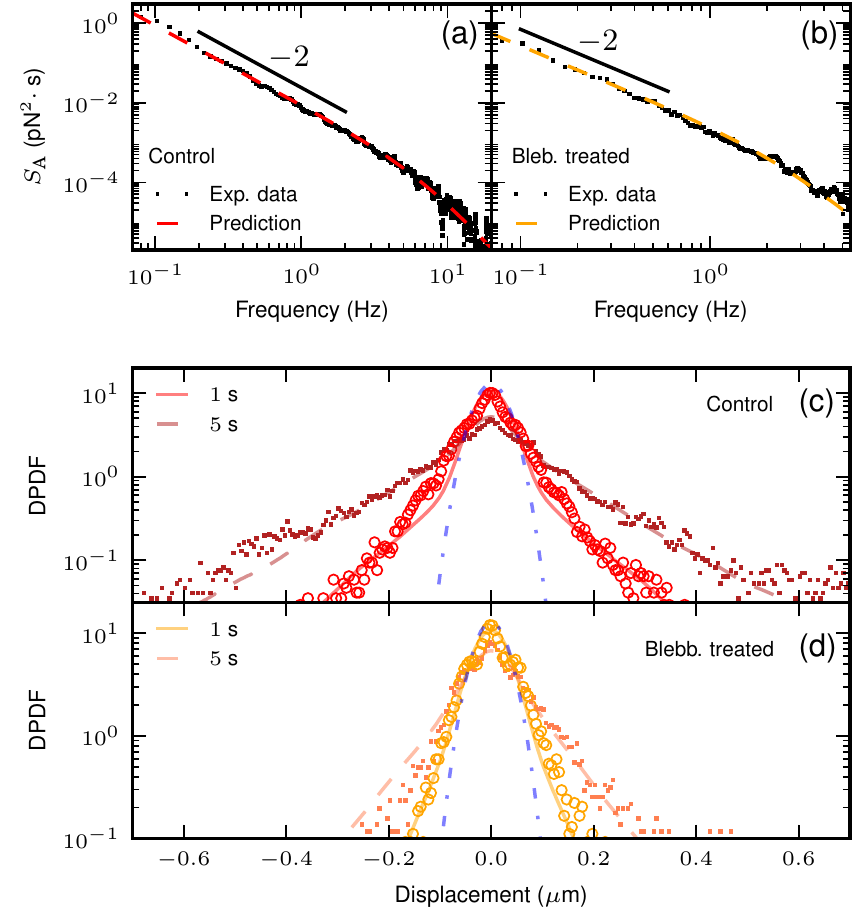}
\caption{\label{fig:pdf} (Top)~Active force spectrum $S_\text{\tiny
    A}$ as a function of frequency measured with tracers of radius
  $a=250$~nm in (a)~control , and (b)~blebbistatin treated cells. The
  experimental data are in black $\cdot\,$, and the dashed lines
  correspond to Eq.~\eqref{eq:spectrum} with the parameter values
  deduced from the best fit of the MSD data.  (Bottom)~Probability
  distribution function of the tracer displacement (DPDF) at two
  different times: $1$ ($\circ$), and $5$~s ($\cdot$). The DPDF is
  measured with tracers of radius $a=250$~nm in (c)~control, and
  (b)~blebbistatin treated cells. We present the corresponding results
  from numerical simulations of Eq.~\eqref{eq-model} in solid
  and dashed lines, respectively. The blue dot dashed line is the
  corresponding equilibrium Gaussian. The parameter values are the
  same for the two lag times:
  (b)~$\{\Ta/T,\tau,\tau_0,k,\g\}=\{2.8\times10^{-3},0.16\,\text{s},2.5\,\text{s},8.5\,\text{pN/$\mu$m},
  4.7\times10^{-3}\,\text{pN}\cdot\text{s}/\mu\text{m}\}$,
  (c)~$\{\Ta/T,\tau,\tau_0,k,\g\}=\{9\times10^{-4},0.39\,\text{s},2.8\,\text{s},8.2\,\text{pN/$\mu$m},
  4.7\times10^{-3}\,\text{pN}\cdot\text{s}/\mu\text{m}\}$.}
\end{figure}

To characterize the properties of the active force, we focus on the
power spectrum of the stress fluctuations, {\it i.e.} the Fourier
transform of the time correlation function
$\avg{\fa(0)\fa(t)}$~\cite{Mizuno,gallet09,Lau}.  We extract the power
spectrum of the overall force $\fa+\xi$ as the power spectrum of the
position times $(6\pi a \abs{G^*})^2$~\cite{Lau}. Provided that the
ATP depleted condition is in an equilibrium state, the active force
$\fa$ is negligible in these cells and the overall force reduces to
the $\xi$, thus providing a direct measurement of the thermal force
spectrum. We remove this equilibrium contribution to the overall
spectrum to deduce the active force spectrum in the two other
conditions. We observe a $1/\omega^2$ behavior at low frequency as
already accounted for on general grounds~\cite{Lau,Loi,Mizuno,Fakhri},
and the large frequency curvature hints a crossover to another power
law (Figs.~\ref{fig:pdf}(a)-(b)). Our analytic prediction for the
active force spectrum reads
\begin{equation}
\label{eq:spectrum} S_\text{\tiny A}(\omega)=\f{1}{\pnt{\omega
    \td}^2}\f{2\gamma\kb\Ta}{1+\pnt{\omega \tau}^2} \,.
\end{equation}
It combines properties of the network and parameters characterizing
the active force, since the effect of nonequilibrium activity on the
tracer is mediated by the network within our model.  We recover the
divergence as $1/\omega^{2}$ at low frequency, and we predict a power
law behavior $1/\omega^4$ at high frequency, the crossover between the
two regimes appearing at $1/\tau$. We compare our prediction with the
experimental data by using the best fit parameters estimated from the
MSD data. Without any free parameter, we reproduce the measured
spectra (Figs.~\ref{fig:pdf}(a)-(b)). This result is a strong support
for our model, in which $\Ta$ not only quantifies the long time
diffusion coefficient of the tracers, it is also related to the
typical amplitude of the fluctuations generated by the nonequilibrium
active force.  The study of the high frequency spectrum calls for new
experiments as it would confirm the validity of our phenomenological
picture.

To study in more details the properties of the active force, we
analyze the probability distribution function of the tracer
displacement (DPDF). It exhibits a Gaussian behavior at short and long
times. In the intermediate regime, we observe a central Gaussian part
which matches our equilibrium prediction in the absence of activity,
and exponential tails accounting for directed motion events consistent
with previous observations in synthetic active
gels~\cite{toyota}. Within our model, the non-Gaussian behavior of the
DPDF is a direct and unique consequence of the non-Gaussianity of the
active force. We ran numerical simulations of the dynamics in
Eq.~\eqref{eq-model} to reproduce the time evolution of the DPDF. We
set the different parameter values to the one estimated previously,
letting us with only one free parameter: the average quiescence time
$\tau_0$. It quantifies the average time between two successive
directed motion events, thus controlling the relative importance of
the exponential tails with respect to the Gaussian central part. We
adjust this parameter by matching the exponential tails observed at
different times.

With a fixed $\tau_0$ value, we manage to reproduce the evolution in
time of the whole experimental DPDF. This shows that the specific form
we choose for the active process is sufficient to reproduce not only
the MSD and force spectrum data, but also to account quantitatively
for the dynamic non-Gaussian properties of the distribution
(Figs.~\ref{fig:pdf}(c)-(d)). We estimate $\tau_0=\{2.5,2.8\}$~s in
control and blebbistatin treated cells, respectively. The extracted
values are very similar for the two conditions, showing that the
addition of blebbistatin does not affect the typical time over which
the tracers are only subjected to thermal fluctuations. It suggests
that this time scale is related to the recovery of the network
following a large reorganization, thus being barely independent of the
activity of the nonequilibrium processes. Notice that the
corresponding duty ratio $p_\text{on}=\tau/(\tau+\tau_0)$ is smaller
in control than in the blebbistatin treated cells:
$p_\text{on}=\{6,15\}\%$, respectively. It is a quantitative evidence
that the exponential tails are more pronounced in the control
condition, namely the proportion of directed motion events is
increased.  We deduce the value of the typical active burst amplitude:
$v=\{0.86,0.22\}$~$\mu$m/s in control and blebbistatin treated cells
for $a=250$~nm, which are compatible with velocity scales observed
in~\cite{guo14}.

Microrheology methods have become a standard technique to explore
cellular activity in living organisms~\cite{guocell}.  In this work,
we introduce a new model for characterizing the motion of a tracer in
a living cell. This model explicitly accounts for the elastic behavior
of the cytoskeletal network and successfully combines it with a
description of the cellular {\em active force}---a well defined
non-Gaussian colored process. By analyzing the MSD data, we quantify
two essential features of this force: its strength, and the typical
time scale over which it is felt. Our model goes beyond previous
modeling which treated the nonequilibrium activity as a random noise
with unprescribed characteristics~\cite{Lau}. In a previous work,
activity was modeled as a trichotomous noise acting directly on the
particle~\cite{Nir}, whereas such activity is mediated by the
surrounding network within our new proposal. The present model
combines the short time confined behavior with a long time free
diffusion which is driven by the active force, and recovers all the
main experimental results. The model applies as long as we are in the
regime of simple viscoelastic behavior. Dressing our model with a more
realistic rheology, \textit{e.g.} with a power law behavior for the
complex modulus, usually observed in cell rheology~\cite{koenderink},
is conceptually straightforward as a future elaboration of the
model. Further generalization of our model could be used to
  describe active fluctuations in other non-equilibrium (living or
  mechanically driven) systems that exhibits similar
  behavior~\cite{weeks,toyota,fodor}.

  We acknowledge several useful discussions with Julien Tailleur and
  Fran\c{c}ois Gallet. N.S.G. gratefully acknowledges funding from the
  Israel Science Foundation (grant no. 580/12).

\bibliographystyle{eplbib}
\bibliography{paper-final}

\end{document}